\documentclass[a4paper]{jpconf}
\usepackage{graphicx}
\begin{document}

\title{The Double Chooz reactor neutrino experiment}

\author{In{\'e}s Gil Botella \footnote{On behalf of the Double Chooz collaboration}}

\address{CIEMAT, Basic Research Department, Avenida Complutense, 22, 28040 Madrid, Spain}

\ead{ines.gil@ciemat.es}

\begin{abstract}
The Double Chooz reactor neutrino experiment will be the next detector to search for a non vanishing $\theta_{13}$ mixing angle with unprecedented sensitivity, which might open the way to unveiling CP violation in the leptonic sector. The measurement of this angle will be based in a precise comparison of the antineutrino spectrum at two identical detectors located at different distances from the Chooz nuclear reactor cores in France. Double Chooz is particularly attractive because of its capability to measure $sin^2(2\theta_{13})$ to 3$\sigma$ if $sin^2(2\theta_{13})$ $>$ 0.05 or to exclude $sin^2(2\theta_{13})$ down to 0.03 at 90\% C.L. for $\Delta$m$^2$ = 2.5 $\times$ 10$^{-3}$ eV$^2$ in three years of data taking with both detectors. The construction of the far detector starts in 2008 and the first neutrino results are expected in 2009. The current status of the experiment, its physics potential and design and expected performance of the detector are reviewed. 
\end{abstract}

\section{Physics motivations and goals}

The neutrino oscillation phenomenon has been clearly established by the study of solar, atmospheric, reactor and beam neutrinos. The PMNS mixing matrix relates the three neutrino mass eigenstates to the three flavor eigenstates. This can be parametrized by three mixing angles $\theta_{ij}$ and one CP violating phase $\delta_{CP}$ (if neutrinos are Dirac particles). During the last years, tremendous progress has been achieved in the experimental field trying to measure the values of $\theta_{ij}$ and the two squared mass differences $\Delta m^{2}_{ij} = m^{2}_{i} - m^{2}_{j}$ which govern the oscillation probabilities. The two mass differences and the $\theta_{12}$ and $\theta_{23}$ mixing angles have been measured with good precision \cite{GonzalezGarcia:2007ib}. However, the $\theta_{13}$ angle, the sign of $\Delta m^{2}_{31}$ and the $\delta_{CP}$ phase are still unknown.

In particular, only an upper limit on the value of $\theta_{13}$ has been established indicating that the angle is very small compared to the other mixing angles. A three-flavor global analysis of the existing data \cite{Schwetz:2006dh} provides a constraint on $\theta_{13}$ being $sin^2\theta_{13}$ $<$ 0.02 at 90\% C.L. This limit is essentially dominated by the result obtained by the CHOOZ reactor experiment \cite{Apollonio:2002gd}. The measurement of this angle is important not only for the final understanding of neutrino oscillations but because it determines the possibilities to observe CP violation in the leptonic sector with the forthcoming neutrino experiments.

Reactor neutrino experiments are able to provide a clean measurement of $\theta_{13}$ as they do not suffer, unlike accelerator experiments, from degeneracies and correlations between different oscillation parameters. They will look for the disappearance of electron antineutrinos produced in nuclear reactors with energies extending up to 10 MeV over distances of the order of kilometers (short baselines) to maximize the disappearance probability. This also prevents reactor experiment measurements to be affected by matter effects.

Reactor antineutrinos are detected through the inverse beta decay $\bar{\nu_e}$ + p $\to$ n + $e^+$, giving a prompt signal due to the $e^+$ annihilation and a delayed signal ($\Delta$t $\sim$ 30 $\mu$s) from the neutron capture. Many liquid scintillator $\bar\nu$ experiments use scintillator loaded with Gadolinium in their fiducial volume because of its large neutron capture cross section and high total $\gamma$ yield of 7-8 MeV.

\section{The Double Chooz concept}

Some of the largest systematic uncertainties of the CHOOZ experiment are related to the accuracy to which the original neutrino flux and spectrum are known. In order to improve the CHOOZ sensitivity, a relative comparison between two or more identical detectors located at different distances from the power plant is required. The first one located at few hundred meters from the nuclear cores monitors the neutrino flux and spectrum before neutrinos oscillate. The second detector located 1-2 km away from the cores searches for a departure from the overall 1/L$^2$ behaviour of the neutrino energy spectrum. At the same time, the statistical error can also be reduced by increasing the exposure and the fiducial volume of the detector.

The Double Chooz experiment \cite{Ardellier:2006mn} will improve our knowledge on the $\theta_{13}$ mixing angle within a competitive time scale and for a modest cost. It will be installed in the Chooz-B nuclear power plant in the Northeast of France. The far detector will be located at 1050 m distance from the cores in the same laboratory used by the CHOOZ experiment. It provides a quickly prepared and well-shielded (300 m.w.e.) site with near-maximal oscillation effect. A second identical detector (near detector) will be installed at $\sim$300--400 m away from the cores to cancel the lack of knowledge of the neutrino spectrum and reduce the systematic errors related to the detector. Since no natural hills or underground cavity already exist, a tunnel will be excavated (overburden 120 m.w.e.) and a near lab will be equipped.

\section{Experimental design}

\begin{figure}[h]
\includegraphics[width=9cm]{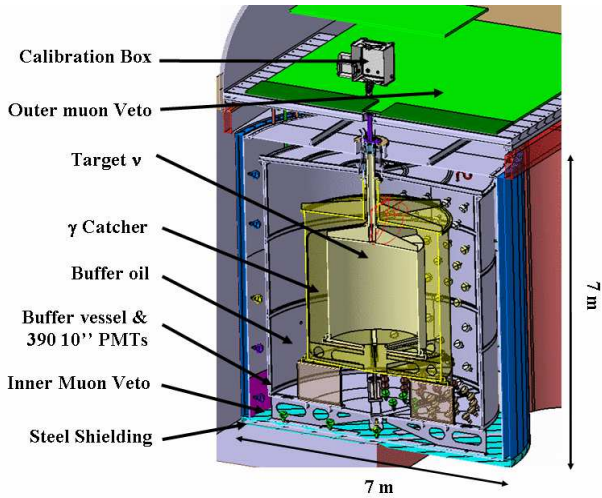}
\begin{minipage}[b]{6cm}
\caption{\label{detector}The Double Chooz detector design.}
\end{minipage}
\end{figure}

The CHOOZ detector design can also be optimized in order to reduce backgrounds. The Double Chooz detectors (see figure \ref{detector}) consist of concentric cylinders and an outer plastic scintillator muon veto. The innermost volume (``target'') contains about 10 tons of Gd-loaded liquid scintillator ($\sim$0.1\% Gd) within a transparent acrylic vessel. It is surrounded by a 55 cm thick layer of unloaded scintillator (``$\gamma$-catcher'') contained in a second acrylic vessel. This scintillating volume is necessary to fully contain the energy deposition of gamma rays from the neutron capture on Gd as well as the positron annihilation gamma rays inside the central region. It also improves the rejection of the fast neutron background. 
Surrounding the $\gamma$-catcher, a 105 cm thick region contains non-scintillating oil inside a stainless steel ``buffer'' vessel. This volume reduces by 2 orders of magnitude with respect to CHOOZ the level of accidental backgrounds coming mainly from the radioactivity of the photomultiplier tubes (PMTs). 390 10'' PMTs are installed on the inner wall and lids of the tank to collect the light from the central scintillating volumes, providing about 13\% photocathode coverage. The central detector is encapsulated within a ``inner muon veto'' shield, 50 cm thick, filled with scintillating organic liquid and instrumented with 78 8'' PMTs. It allows the identification of muons passing near the active detector that can create spallation neutrons and backgrounds coming from outside. Because of space constraint, the 70 cm sand shielding of CHOOZ is replaced by a 15 cm iron layer to protect the detector from rock radioactivity and to increase the target volume. An ``outer muon veto'' covers the top of the main system and provides additional rejection power for cosmic-induced events. It can be used for constant mutual efficiency monitoring with the inner veto.

The near and far detectors will be identical inside the PMT support structure, allowing a relative normalization error of 0.6\% or less, to be compared with a 2.7\% systematic error of the CHOOZ experiment. 

\section{Systematic errors and backgrounds}

Many systematic uncertainties that affected CHOOZ and all previous single-baseline reactor neutrino experiments are greatly reduced by having both near and far detectors. Table \ref{sys} summarizes the systematic uncertainties in the measurement of the antineutrino flux comparing both CHOOZ and Double Chooz detectors.

\begin{table}[h]
\caption{\label{sys}Systematic uncertainties in CHOOZ and Double Chooz experiments.} 
\begin{center}
\begin{tabular}{lll}
\br                              
  &  CHOOZ  &  Double Chooz\\
\mr
Reactor fuel cross section & 1.9\% & -- \\
Reactor power & 0.7\% & -- \\
Energy per fission & 0.6\% & -- \\
Number of protons & 0.8\% & 0.2\% \\
Detection efficiency & 1.5\% & 0.5\% \\
\br
\end{tabular}
\end{center}
\end{table}

Though an uncertainty from the neutrino contribution of spent fuel pools remains, it is negligible for Double Chooz. The neutrino rates are proportional to the number of free protons inside the target volumes, which thus has to be experimentally determined with a precision of 0.2\%. This constitutes one of the major improvements with respect to CHOOZ. In addition, a comprehensive calibration system consisting of radioactive sources deployed in different detector regions, laser light flashes and LED pulses, will be enforced to correct the unavoidable differences between the two detector responses. The optimization of the Double Chooz detector design allows to simplify the analysis and to reduce the detection efficiency systematic errors up to 0.5\% while keeping high statistics.

The main sources of background in the detector can be identified as accidentals, due to the random coincidence of an environmental gamma (positron-like signal) and a neutron-like event within the neutrino time window; fast neutrons produced by cosmic muons, which can produce proton-recoils in the target (misidentified as e$^+$) and then be captured; and long-lived cosmogenic radioisotopes ($^9$Li, $^8$He) with significant branching for $\beta$-decay followed by neutron emission. 

The selection of high pure materials for detector construction and passive shielding around the active region provide an efficient protection against accidentals. Furthermore, they can be measured {\it in situ} even with reactors on. The inner and outer veto systems and the inner detector muon electronics are designed to address the correlated backgrounds.

\section{Expected sensitivity}

\begin{figure}[h]
\begin{center}
\includegraphics[width=10cm]{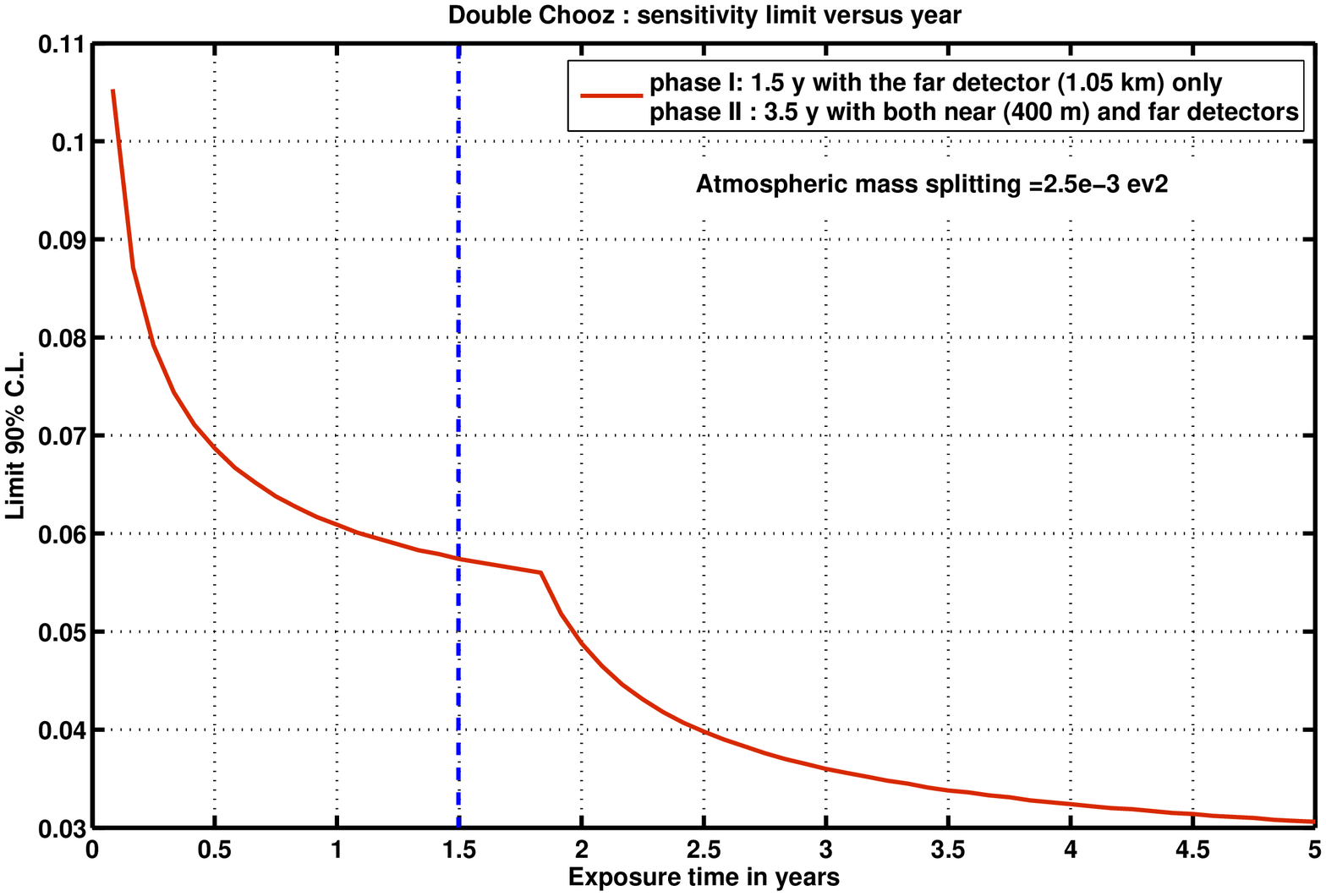}
\end{center}
\caption{\label{sensit}Double Chooz expected sensitivity limit (90\% C.L.) to $sin^2(2\theta_{13})$ as a function of time, assuming that the near detector is built 1.5 years after the far detector.}
\end{figure}

The Double Chooz experiment will make a fundamental contribution to the determination of the $\theta_{13}$ mixing angle within an unrivalled time scale. In a first phase, only with the far detector, Double Chooz will be sensitive to $sin^2(2\theta_{13})$ $>$ 0.06 after 1.5 years of data taking (see figure \ref{sensit}), a factor 3 better than CHOOZ. In a second phase, with both near- and far- detectors running simultaneously, Double Chooz will explore $sin^2(2\theta_{13})$ $>$ 0.03 after 3 years of operation.

\section{Current status and conclusions}

An intense R\&D work has been carried out by the Double Chooz collaboration to validate the robustness of the detector concept and the feasibility of the sensitivity goals. The experiment is now in the construction phase of the far detector, starting the integration in the existing underground laboratory early 2008. First neutrino results are expected for 2009. The near lab site will be available by the end of 2009 to accommodate the second detector.

Double Chooz will be able to measure $sin^2(2\theta_{13})$ to 3$\sigma$ if $sin^2(2\theta_{13})$ $>$ 0.05. Otherwise, it will exclude the mixing angle down to $sin^2(2\theta_{13})$ $>$ 0.03 at 90\% C.L. after 3 years of operation with both detectors. This will represent an increase of about a factor 7 compared to CHOOZ and will open the way for a new level of accuracy in reactor neutrino experiments. The information gained with Double Chooz will complement future results with accelerator experiments, affected by degeneracy problems, helping to better constrain the last undetermined mixing parameters.

\section*{References}

\end{document}